\documentstyle[aps,epsf,prl,multicol]{revtex}
\draft

\begin{document}
\title{Path instabilities of  air bubbles rising  in clean  water } 
\author{Mingming Wu}
\address{Department of Physics, Occidental College, Los Angeles, CA 90041, USA}
\author{Morteza Gharib}
\address{ Graduate Aeronautical Laboratories, California Institute of Technology, Pasadena, CA 91125, USA} 
 \date{\today} \maketitle
\begin{abstract}
Experiments are conducted to study the  path and shape of  single  air bubbles  (diameter range   $0.10- 0.20cm$)  rising freely in clean   water.  The experimental  results  demonstrate  that 
the bubble   shape  has  a bistable state,   {\it i. e.}  the bubble  chooses to be  in   spherical or   ellipsoidal  shape  depending on its   generation mechanism.   The path of a  spherical/ellipsoidal  bubble   is  found to change  from a  straight path    to a zigzag/spiral  
 path  via a supercritical/subcritical bifurcation when the Reynolds number  of the  bubble
 exceeds  a threshold.   


\end{abstract}

\begin{multicols}{2}

The spiral or  zigzag motion of  air bubbles rising freely  in a fluid medium have been observed  in various experiments.\cite{HM53,PGS56,HS57,TH77,LP97}   Extensive work has been done to  determine the criteria for the onset of  path instability.\cite{HS57,TH77,RL84,DM89}   
Simple as the matter appears to be, it turns out that  the path of the bubble itself is very sensitive to the experimental setup near the instability point, especially to  the turbulence level in the fluid medium due to  the background noise and the contaminations of  the fluids.  As a consequence, no consistent stability criteria can be found  
in the current literature. The exact nature of the path instability is yet to be explored.


The focus of our investigations is on the understanding of the underlying mechanism of the path instability. 
Recent work by Kelley and Wu\cite{KW97}  on path instabilities of a penny-shaped bubble rising in a Hele-Shaw cell (2-dimensional case) demonstrated that the path of a  bubble was changed  from a straight path to a zigzag path when the 
Reynolds number exceeded  a critical value. Colored dye  visualization experiments  showed that such instability was a consequence of vortex shedding in the wake of the bubble,  a reminiscence of vortex shedding in the wake of a solid cylinder.  In the 3-dimensional case where bubbles rise freely in fluids,  some intriguing work  has been done   by Lunde and Perkins\cite{LP97}  where the wake structures of the bubbles   rising  in tap water  were studied  using colored dye visualization technique in the Reynolds number range of 600-1700.   The experimental results demonstrated a clear connection between the wake structures and the lateral motions of the bubbles.  
  
In this letter, we present experimental investigations  on the shape and path of the rising bubbles in clean  water near and above  the path instability point.   
The main part of the experimental apparatus   is  the plexiglass water tank  with dimension 
$6''\times 6'' \times 24''$.  At the center of the bottom plate, a  specially designed fitting  is   mounted  for the hypodermic needle to go through. The bubble  is   released through the  hypodermic needle.  One camera is   used to image the shape and size  of the bubble, and it is   mounted close to the position where the bubble is   released.  The camera takes   an image of 640$\times 480 $ pixels with a viewing window of $1.25cm\times 0.936cm$. 
The second  camera  is   a 
specially designed 3-D imaging system,\cite{WG92}  and it   is   used to map out the (x, y, z) coordinates of the bubble. The camera obtains the third dimension $z$  (direction of rising bubble) using a  quantitative defocusing mechanism. A typical viewing window of the 3-D imaging system  is  
$1.5cm\times 1.5cm \times 20cm$.  
 Extreme cautions   were   taken to keep the tank  as clean as possible.  Doubly deionized and distilled water  was   used. Large bubbles  were   driven  through the tank  for  removing the surface-active contaminants prior to each experimental run.   The temperature of the 
water tank  was $24.5\pm0.3 ^o$C during  the experimental runs.  The kinetic viscosity 
$\nu = 0.00901$ cm$^2$/s and  surface tension $\sigma = 72.1$ dyn/cm are  given by  Ref. \cite{CRC85}.    

\narrowtext
\begin{figure}[t]
\epsfxsize=3.25in \centerline{\epsffile{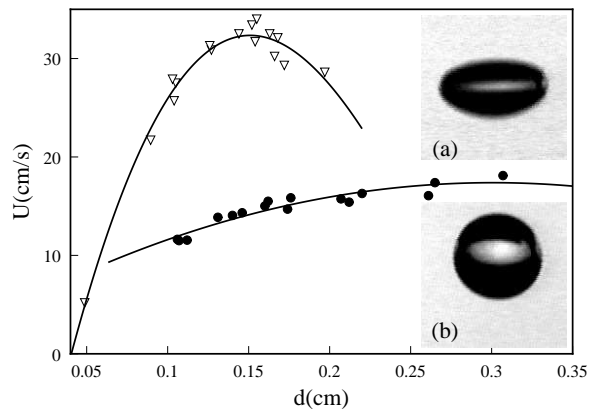}}
\vskip 0.1in
\caption{ Terminal   velocity    $U$ versus   diameter $d$ 
of   $\bigtriangledown$: ellipsoidal bubbles and $\bullet$: nearly spherical bubbles. 
Inset (a) is an image of  an
ellipsoidal bubble  with   $d =  0.172cm$,  inset (b)  is an image of a nearly spherical bubble 
with  $d = 0.174cm$. }
\end{figure}

Two different  bubble generation methods have been  used. The first one is   to attach the 
hypodermic needle 

\narrowtext
\begin{figure}[t]
\epsfxsize=3.5in \centerline{\epsffile{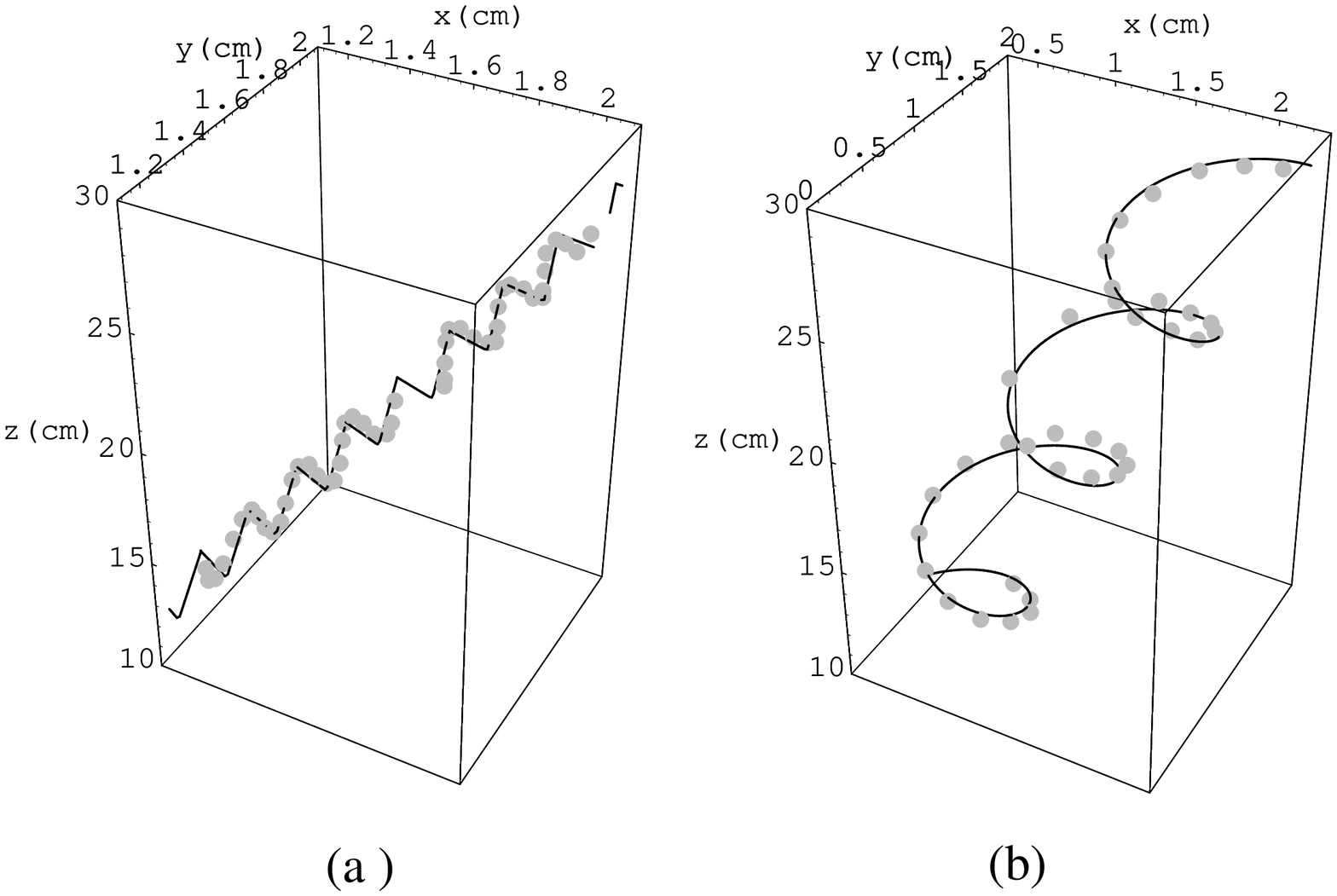}}
\vskip 0.1in
\caption{  (a) Zigzag path of a nearly spherical bubble at $Re = 269$.  Solid line is a fit to a zigzag function.  (b) Spiral path of an ellipsoidal bubble at $Re = 640$. Solid line is a fit to a spiral function.  In both (a) and (b), the dotted lines are from experiments.  }
\end{figure}

\noindent
directly to a syringe filled with air. Push the syringe gently until a bubble is  formed at the tip of the needle and then is    pinched  off from the needle.
The size of the bubble  depends  on  the 
 the  inner diameter  of the needle and the shape of the needle tip.  Using this pinch off method,  we have consistently generated bubbles with ellipsoidal shape.  The   aspect ratios  (long axis  versus short axis ) of the bubbles  are   between   1.12--1.89 for bubbles of    diameter  range of  $0.10cm- 0.20cm$.   The diameter here is   defined as $(6V/\pi)^{1/3}$, where $V$ is the volume of the bubble.  $V$ is   obtained using the image taken by  the CCD camera.  Fig. 1 shows a  typical image of an ellipsoidal bubble.      For  the second  bubble generation method,  the  hypodermic needle is attached  to a three way valve, of which one way is connected to  a syringe filled with water and the other to  a  syringe pump filled with air.     The hypodermic needle has  a specially designed  capillary tube with a flat top.   The inner diameter of the tube is $0.121cm$  and the  length of the tube is $3''$.
 To generate a bubble, a desired volume of air
  is pushed into  the lower end of the capillary tube by the  syringe pump, and then  the  direction of  a three way valve is switched so that the bubble can be gently pushed out of the capillary tube  by the syringe filled with water.  
Using this  gentle push method, we were able to obtain consistently bubbles of nearly  spherical shape. The aspect ratios  are   between 1.00 -1.07 for bubbles in the   diameter range of $0.10-0.20cm$. A typical spherical bubble image is shown in Fig. 1.  It needs to be noted here that the diameter range in  which the bubble has a bistable shape state extends  beyond $0.10-0.20cm$.  

The velocities of the bubbles of  various  diameters are shown  in Fig. 1. It  demonstrates  that for the same diameter, the ellipsoidal bubble moves much faster than the spherical bubble. 
The measured velocities  of the ellipsoidal  bubbles  are  consistent  with those obtained by

\narrowtext
\begin{figure}[t]
\epsfxsize=3.25in \centerline{\epsffile{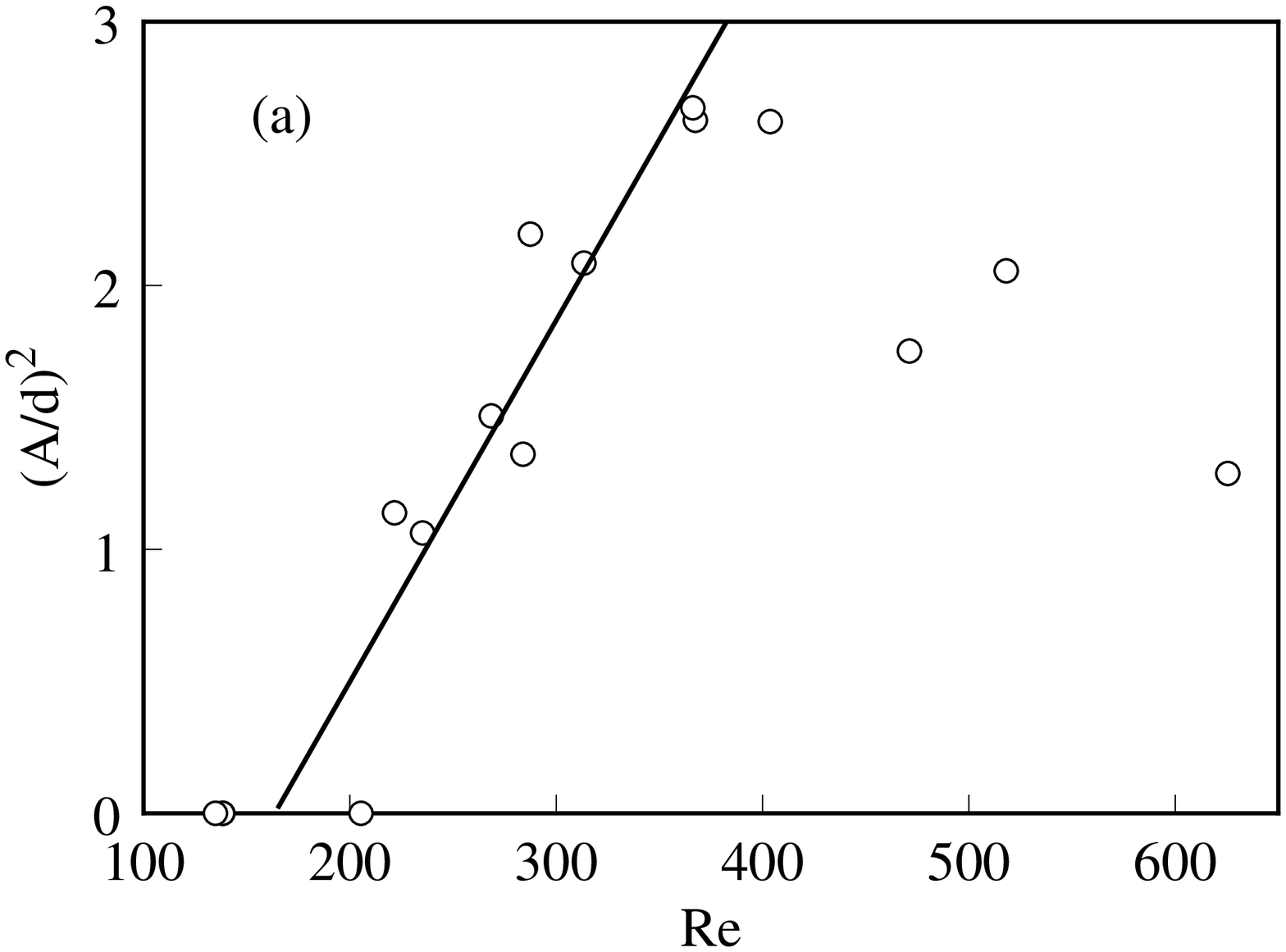}}
\vskip 0.1in
\epsfxsize=3.25in \centerline{\epsffile{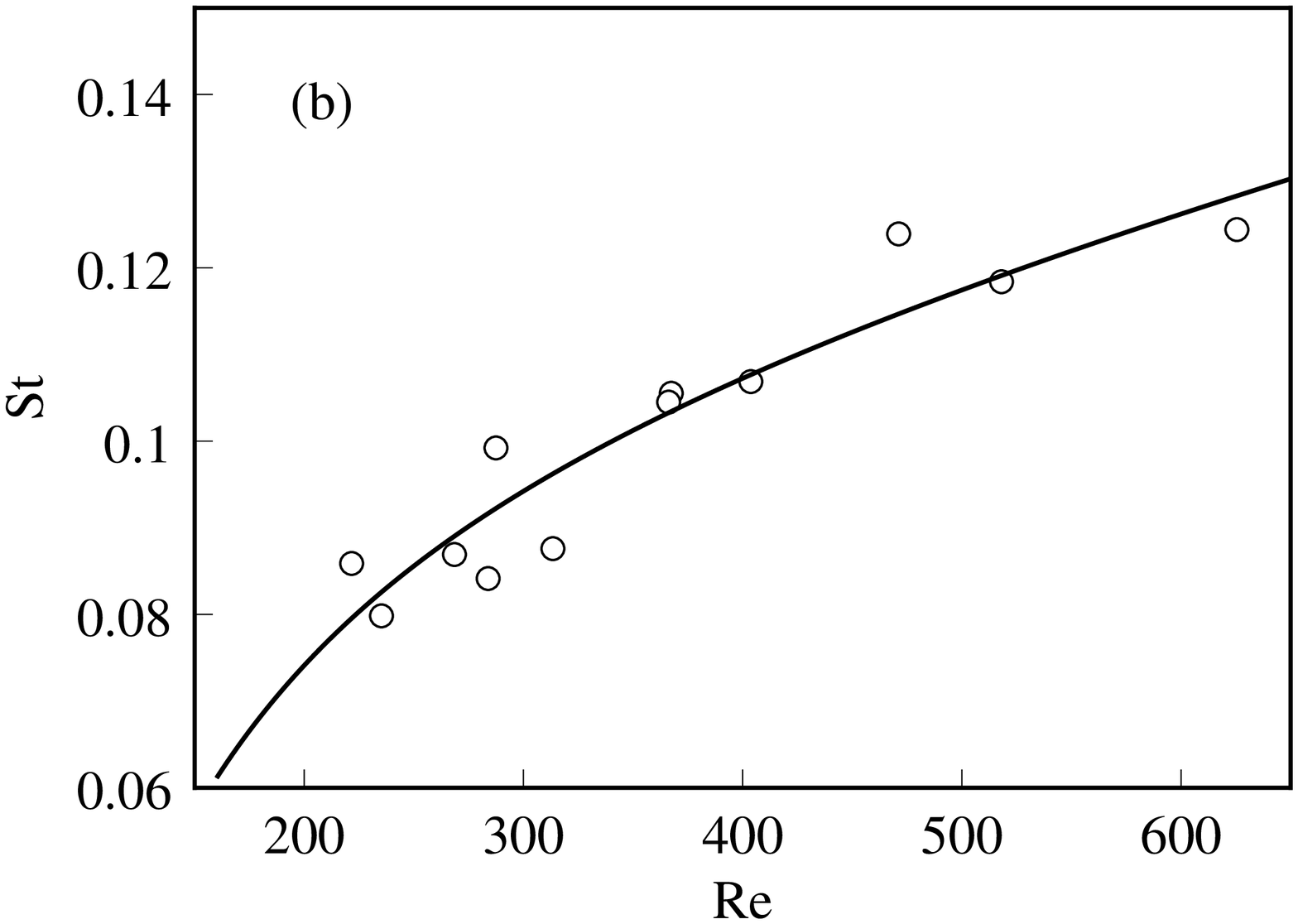}}

\caption{   (a)  Curves of  dimensionless amplitude squared   versus  Reynolds number for nearly spherical bubbles. Solid line is a  fit to  a linear function.  (b) Curves of  Strouhal  versus Reynolds number. Solid line is a  fit to $A/R +B+CR$, $A = -8.50, B = 0.105,  C = 5.94\times 10^{-5}$. Dotted lines are from experiments. }
\end{figure}

\noindent
Maxworthy {\it et.  al} using clean  water. \cite{MGKD96}  Note that pinch off method was used 
in their experiment.\cite{MAX97}
 
For nearly  spherical bubble, the straight path of the bubble   is changed  to a zigzag path 
as the Reynolds number  $Re$ exceeds   a critical value $Rec$.  Here Reynolds
number $Re $ is defined as $ U d /\nu$, where $U$ is the bubble velocity, $d$ is the bubble diameter, and $\nu$ is the kinetic viscosity of the fluid.  Typical planar zigzag path of the bubble is shown in Fig. (2a).  A zigzag function is used to fit the zigzag path, and  the amplitude $A$ and the frequency $f$ of the zigzag path is obtained from the fitted parameters.   In Fig. (3a), the dimensionless  amplitude squared of the zigzag path  is plotted as a function of Reynolds number. As seen that the amplitude squared is linearly related to the Reynolds number near the onset of the path instability,  a signature of supercritical bifurcation.  The linear extrapolation of the line gives the critical Reynolds number of $Rec  = 157\pm 10$.  The Strouhal curve of the zigzag path  is shown in Fig. (3b). The Strouhal number here is defined as $St = fd/U$. 
Colored dye visualization experi-

\narrowtext
\begin{figure}[t]
\epsfxsize=3.25in \centerline{\epsffile{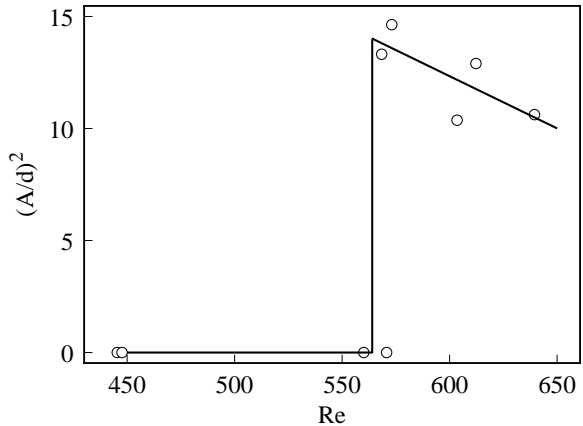}}
\vskip 0.1in
\caption{  Curves of  dimensionless amplitude squared   versus  Reynolds number for ellipsoidal bubbles. Solid lines are guidelines. }
\end{figure}

\noindent 
 ments   were carried out and the results were 
 consistent with those of   Lunde and Perkins\cite{LP97}, where hairpin vortices were observed to shed periodically from the alternate rear sides of the zigzag bubble. 
This indicates that the zigzag path instability is   caused by vortex shedding in the wake of the bubble, in a similar way as in the 2-D situation.\cite{KW97}


The path instabilities of the ellipsoidal bubble differs from those of spherical bubble qualitatively.   The straight path of the ellipsodal bubble changes to a spiral path as shown in Fig. 2b  when the Reynolds number  exceeds a critical value.  The measured spiral path is fitted to a spiral function, where the amplitude $A$ and frequency of the spiral path are obtained from the fitted parameters. The curve of dimensionless amplitude squared versus Reynolds number (Fig. 4) shows  that  the spiral path instability occurs at   $Rec = 564\pm 10$  via  a  subcritical bifurcation. The Strouhal number at the onset is $\sim 0.02$.    Colored dye  
visualization experiments on the wake structures behind spiralling bubbles by 
Lunde and Perskin\cite{LP97}  revealed  two continuous vortex filaments. 
 At present, the exact cause for the spiral instability  is yet to be explored.   We do notice that the short axis of the ellipsoidal bubble is always aligned with the direction of the bubble motion, it is likely that the ellipsoidal bubble spins around its short axis as it spirals up.

\noindent

In summary,   We find that within  a certain diameter range, the bubble  shape  has a bistable state, it   can be 
either in  spherical  or  ellipsoidal shape depending on its generation mechanism. The 
spherical bubble undergoes  a zigzag path instability as the Reynolds number exceeds $157\pm 10$ via a supercritical bifurcation. The ellipsoidal bubble undergoes a spiral path instability as the Reynolds number exceeds $564\pm 10$ via a subcritical bifurcation. 

Wu would like to thank Prof. Maxworthy for an insightful   discussion on the subject. 
Wu also would  like to thank members of Gharib's group who offered generous help during her  summer stay at  Caltech.  This work is supported by the Office of Naval Research (URI research grant   N00014-92-8-1618)  and the   Petroleum Research  Fund (ACS-PRF\#  32904-GB9).




\end{multicols}
\end{document}